# Kinetics of atomic ordering in metal-doped graphene

Radchenko Taras M., Tatarenko Valentyn A.

*Dept. of Solid State Theory, Institute for Metal Physics, N.A.S.U.,
36 Acad. Vernadsky Boulevard, UA-03680 Kyiv-142, Ukraine*

Corresponding author (Radchenko T.M.). Tel. +380 44 424 12 21; fax: +380 44 424 25 61.
*E-mail address:* taras.radchenko@gmail.com (Radchenko T.M.)

Possible stably-ordered substitutional structures based on a graphene-type crystal lattice are considered. A kinetic model of atomic ordering in metal-doped graphene with stoichiometric (1/8, 1/4, 1/2) and non-stoichiometric compositions is developed. Inasmuch as the intrasublattice and intersublattice 'interchange' ('mixing') energies are competitively different for graphene-based lattice, kinetic curves of the long-range order (LRO) parameters may be nonmonotonic for the structures described by two or three LRO parameters.

## 1. Introduction

Graphene—a one-atom-thick layer modification of carbon recently discovered in the free state [1,2]—is a hot-topic object in both materials science and condensed matter physics, where it is a popular model system for investigations. Graphene is the basic structural element of graphite lattice. Crystal lattice of graphene is a nanoscale structure (so-called two-dimensional carbon), where atoms are distributed over the vertexes of regular hexagons, as shown in Fig. 1a. Due to a high mechanical strength, hardness, heat conductivity [3], and electrical conductivity, graphene is a perspective material for a wide application in the different fields: from nanoelectronics (graphene will be its basis [4,5], substituting silicon in the integrated circuit chips) to coating of airliner fuselage.

Probably, graphene doping with metal (*Me*) atoms may improve some of its physical properties (for a wider range of application). Particularly, the doping with a metal changes the band structure (which strongly depends on atomic order) and, consequently, improves an electrical conductivity of graphene [6–21].

This work is focused on the construction of both statistical-thermodynamic and kinetic models for long-range atomic order in a metal-doped graphene, i.e. in a two-dimensional substitutional solid solution C–*Me* based on a graphene-type crystal lattice.

## 2. Model

Since the graphene-type lattice is a two-dimensional 'honeycomb'-crystal lattice consisting of two interpenetrating hexagonal sublattices, as shown in Fig. 1a, its reciprocal lattice is a two-dimensional hexagonal lattice as well (Fig. 1b). Nearest-neighbour distance (between C atoms) in hexagons (Fig. 1a) is $a_0 \approx 0.142$ nm [6]. It is conveniently to consider a graphene-type lattice as consisting of two interpenetrating hexagonal sublattices displaced with respect to each other by the vector $\mathbf{h} = \mathbf{a}_1/3 + 2\mathbf{a}_2/3$, where $\mathbf{a}_1$ and $\mathbf{a}_2$ are the fundamental translation vectors of a lattice along the [1 0] and [0 1] directions, respectively, in the oblique system of coordinates (see Fig. 1a). Lattice translation parameter is $a = |\mathbf{a}_1| = |\mathbf{a}_2| = \sqrt{3}a_0 \cong 0.246$ nm. As shown in Fig. 1a, each *ABCD* primitive unit cell contains two sites. Each lattice-site location can be described by a sum of two vectors: $\mathbf{R} + \mathbf{h}_q = \mathbf{r}$ (Fig. 1a). Vector $\mathbf{R}$ denotes an 'origin' position of the unit cell. Vector $\mathbf{h}_q$ denotes the distance of a given site with respect to the unit-cell 'origin', and $q$ subscript numerates the sublattice ($q = 1, 2$). The radius-vector $\mathbf{R}$ is related to the basis vectors as $\mathbf{R} = n_1\mathbf{a}_1 + n_2\mathbf{a}_2$, where $n_1, n_2$ are integers.

Let us consider possible stably-ordered (super)structures of two-dimensional substitutional C–*Me* solid solution based on the graphene-type lattice with superstructural stoichiometric $C_3Me$, $C_7Me$,

C*Me* compositions (Fig. 2–4). Single-site occupation-probability functions for these (super)structures are obtained using the static-concentration-waves' approach [22] and presented in Table 1, where $P_q(\mathbf{R})$ is the probability to find a *Me* atom at the $(q, \mathbf{R})$ site, i.e. at the site of $q$-th sublattice within the unit cell with 'origin' $\mathbf{R}$, and $\eta_\sigma^\aleph$ ($\sigma = 0$, 1 or 2) are the LRO parameters ($\aleph$ index denotes their total number for a given structure; $\aleph =$ I, II or III).

Interatomic interactions in C–*Me* lattice can be taken into consideration by means of the 'mixing' ('interchange') energies [22]:

$$w_{pq}(\mathbf{R} - \mathbf{R}') \equiv W_{pq}^{CC}(\mathbf{R} - \mathbf{R}') + W_{pq}^{MeMe}(\mathbf{R} - \mathbf{R}') - 2W_{pq}^{CMe}(\mathbf{R} - \mathbf{R}').$$

Here, $p$ and $q$ subscripts number the sublattices, where corresponding atoms can be distributed; $W_{pq}^{CC}(\mathbf{R} - \mathbf{R}')$, $W_{pq}^{MeMe}(\mathbf{R} - \mathbf{R}')$, $W_{pq}^{CMe}(\mathbf{R} - \mathbf{R}')$ are the pair-wise interaction energies of C–C, *Me*–*Me*, C–*Me* pairs of atoms, respectively, located at the sites of $p$-th and $q$-th ($p, q = 1, 2$) sublattices within the unit cells with origins ('zero' sites) at $\mathbf{R}$ and $\mathbf{R}'$ sites.

For a statistical-thermodynamic description of the arbitrary-range interatomic-interactions (i.e. in all coordination shells), it is conveniently to apply the Fourier transformations for the elements of the 'mixing'-energies matrix [22],

$$\|\tilde{w}_{pq}(\mathbf{k})\| \equiv \begin{pmatrix} \tilde{w}_{11}(\mathbf{k}) & \tilde{w}_{12}(\mathbf{k}) \\ \tilde{w}_{12}^*(\mathbf{k}) & \tilde{w}_{11}(\mathbf{k}) \end{pmatrix}, \text{ where } \tilde{w}_{pq}(\mathbf{k}) \equiv \sum_{\mathbf{R}} w_{pq}(\mathbf{R} - \mathbf{R}') e^{-i\mathbf{k}\cdot(\mathbf{R}-\mathbf{R}')}.$$

Here, $\mathbf{k}$ is a wave vector of a two-dimensional reciprocal space (Fig. 1b), which 'generates' corresponding (super)structure; $\tilde{w}_{12}^*(\mathbf{k})$ is a complex conjugate to $\tilde{w}_{12}(\mathbf{k})$. Writing Hermitian 'mixing'-energy matrix, the symmetry relations, $\tilde{w}_{11}(\mathbf{k}) = \tilde{w}_{22}(\mathbf{k})$ and $\tilde{w}_{21}(\mathbf{k}) = \tilde{w}_{12}^*(\mathbf{k})$, are taken into account.

The 'mixing' energies and corresponding eigenvalues of the 'mixing'-energy matrix $\|\tilde{w}_{pq}(\mathbf{k})\|$,

$$\lambda_1(\mathbf{k}) = \tilde{w}_{11}(\mathbf{k}) + |\tilde{w}_{12}(\mathbf{k})|, \quad \lambda_2(\mathbf{k}) = \tilde{w}_{11}(\mathbf{k}) - |\tilde{w}_{12}(\mathbf{k})|,$$

entering into expressions for the configurational free energy, which can be obtained within the self-consistent field approximation [22], define the statistical thermodynamics and kinetic behaviour of the doped graphene-based structures.

Substitution of the functions from Table 1 into the configurational Helmholtz free-energy functional [22],

$$F \cong \frac{1}{2} \sum_{p,q=1}^{2} \sum_{\mathbf{R},\mathbf{R}'} w_{pq}(\mathbf{R} - \mathbf{R}') P_p(\mathbf{R}) P_q(\mathbf{R}') + k_B T \sum_{q=1}^{2} \sum_{\mathbf{R}'} \left[ P_q(\mathbf{R}') \ln P_q(\mathbf{R}') + (1 - P_q(\mathbf{R}')) \ln (1 - P_q(\mathbf{R}')) \right]$$

($T$ is a temperature, $k_B$—Boltzmann constant), and simple mathematical transformations yield expressions [23] for the configurational free energy (per atom) of the (super)structures, which are 'generated' high-symmetry point wave vectors and stable against the antiphase shifts (Figs. 2–4).

Within the model for the long-range atomic-order kinetics, we consider the case of exchange ('ring') diffusion mechanism [22–35] 'governing' atomic ordering in a two-dimensional binary solution based on the graphene-type lattice. To investigate the ordering kinetics of C and *Me* atoms over the sites of this lattice, a model based on the Onsager-type microdiffusion master equation [22–35] was applied:

$$\frac{dP_p^\alpha(\mathbf{R},t)}{dt} \approx -\frac{1}{k_B T} \sum_{q=1}^{2} \sum_{\mathbf{R}'} \sum_{\beta = C, Me} c_\alpha c_\beta L_{pq}^{\alpha\beta}(\mathbf{R} - \mathbf{R}') \frac{\delta F}{\delta P_q^\beta(\mathbf{R}',t)},$$

where $t$ is a time, $c_\alpha$ ($c_\beta$) is the relative fraction of $\alpha$-kind ($\beta$-kind) atoms, $L_{pq}^{\alpha\beta}(\mathbf{R} - \mathbf{R}')$ is a matrix of kinetic coefficients whose elements represent probabilities of elementary exchange-diffusion jumps of a pair of $\alpha$ and $\beta$ atoms at the $\mathbf{r}$ site of $p$-th sublattice and $\mathbf{r}'$ site of $q$-th sublattice, respectively ($\alpha, \beta =$ C, *Me*).

Applying the conservation condition for the atoms of each kinds, the fact that each lattice site is definitely occupied by one of the atoms composing the binary solution, and the Fourier transformation for the last equation (in details, see [23]), we derived differential equations (see Table 2) for the kinetics of LRO parameter(s) of the (super)structures shown in Figs. 2–4. In Table 2, reduced time $t^*$ is defined by the Onsager-type kinetic coefficients, $t^* \equiv \tilde{L}(\mathbf{k})t$, where $\tilde{L}(\mathbf{k})$ is the Fourier-transform of a certain concentration-dependent combination of $\{L_{pq}^{\alpha\beta}(\mathbf{R}-\mathbf{R}')\}$, and reduced temperature $T^*$ depends on the 'mixing' energies as follows: $T^* = k_B T/|\lambda_2(\mathbf{k})|$.

## 3. Results and conclusions

Curves in Figs. 5, 6 represent numerical solutions of differential kinetics equations for LRO of stoichiometric (Fig. 5; [23]) and nonstoichiometric (Fig. 6) $C_{1-c}Me_c$ structures ($c$ is atomic fraction of doping *Me*-component) at the reduced temperature $T^* = 0.1$.

Slight (Figs. 5a, 6a, 6d) and significant (Figs. 5d, 5e) nonmonotonies of the evolution of the LRO parameters are caused by the competitive difference of intrasublattice and intersublattice 'mixing' energies.

As shown in Figs. 5a–5c, an initial value of the LRO parameter (e.g., $\eta_\sigma^{\aleph}(t^*=0) = 0.1$ or $\eta_\sigma^{\aleph}(t^*=0) = 0.3$) does not affect its end ('equilibrium') value for a given structure. As expected, this value is the same at other equal conditions.

Kinetic curves for $C_{1-c}Me_c$ structures with deviation of $\Delta c = \pm 0.01$ from stoichiometry are represented in Fig. 6. It is interesting that, for equiatomic (C*Me*) (super)structures, decrease and increase of atomic fraction of dopant do not affect the relaxation kinetics of LRO parameter: its instantaneous (and 'equilibrium') values are equal in Figs. 6c, 6f. However, for two other (super)structures ($C_3Me$ and $C_7Me$), decreasing and increasing of the relative concentration of dopant with respect to its stoichiometry affect kinetic process differently. Decrease in this case reduces instantaneous (and 'equilibrium') LRO-parameter values, and increase of the dopant concentration may elevate instantaneous (and 'equilibrium') LRO-parameter values (see Figs. 6a, 6b, 6d, 6e in comparison with Figs. 5a, 5b).

In closing, it is important to note that possible interstitial (super)structures based on a graphene-type lattice will be considered as well in another article.

## Acknowledgement

This work was supported by the Grant of the National Academy of Sciences of Ukraine within the framework of research project for young scientists, which is gratefully acknowledged.

Table 1. Single-site occupation probabilities for dopant *Me*-atoms in graphene-based (super)structures.

| (Super)structure | Probability function |
|---|---|
| $C_3Me$ (Fig. 2a) | $\begin{pmatrix} P_1(\mathbf{R}) \\ P_2(\mathbf{R}) \end{pmatrix} = c\begin{pmatrix} 1 \\ 1 \end{pmatrix} + \frac{1}{4}\eta_2^I\left[\begin{pmatrix} 1 \\ -1 \end{pmatrix}\cos(\pi n_1) + \begin{pmatrix} 1 \\ 1 \end{pmatrix}\cos(\pi n_2) + \begin{pmatrix} 1 \\ -1 \end{pmatrix}\cos(\pi(n_1 - n_2))\right]$ |
| $C_3Me$ (Fig. 2b) | $\begin{pmatrix} P_1(\mathbf{R}) \\ P_2(\mathbf{R}) \end{pmatrix} = c\begin{pmatrix} 1 \\ 1 \end{pmatrix} + \frac{1}{4}\left[\eta_1^{II}\begin{pmatrix} 1 \\ 1 \end{pmatrix}\cos(\pi n_1) + \eta_2^{II}\begin{pmatrix} 1 \\ 1 \end{pmatrix}\cos(\pi n_2) + \eta_1^{II}\begin{pmatrix} 1 \\ 1 \end{pmatrix}\cos(\pi(n_1 - n_2))\right]$, <br><br> $\begin{pmatrix} P_1(\mathbf{R}) \\ P_2(\mathbf{R}) \end{pmatrix} = c\begin{pmatrix} 1 \\ 1 \end{pmatrix} + \frac{1}{4}\left[\eta_1^{II}\begin{pmatrix} 1 \\ 1 \end{pmatrix}\cos(\pi n_1) + \eta_1^{II}\begin{pmatrix} 1 \\ -1 \end{pmatrix}\cos(\pi n_2) + \eta_2^{II}\begin{pmatrix} 1 \\ -1 \end{pmatrix}\cos(\pi(n_1 - n_2))\right]$, <br><br> $\begin{pmatrix} P_1(\mathbf{R}) \\ P_2(\mathbf{R}) \end{pmatrix} = c\begin{pmatrix} 1 \\ 1 \end{pmatrix} + \frac{1}{4}\left[\eta_2^{II}\begin{pmatrix} 1 \\ -1 \end{pmatrix}\cos(\pi n_1) + \eta_1^{II}\begin{pmatrix} 1 \\ -1 \end{pmatrix}\cos(\pi n_2) + \eta_1^{II}\begin{pmatrix} 1 \\ 1 \end{pmatrix}\cos(\pi(n_1 - n_2))\right]$ |
| $C_3Me$ (Fig. 2c) | $\begin{pmatrix} P_1(\mathbf{R}) \\ P_2(\mathbf{R}) \end{pmatrix} = c\begin{pmatrix} 1 \\ 1 \end{pmatrix} + \frac{1}{4}\eta_0^{III}\begin{pmatrix} 1 \\ -1 \end{pmatrix} + \frac{1}{4}\left[\eta_1^{III}\begin{pmatrix} 1 \\ 1 \end{pmatrix} + \eta_2^{III}\begin{pmatrix} 1 \\ -1 \end{pmatrix}\right]\cos(\pi(n_1 - n_2))$, <br><br> $\begin{pmatrix} P_1(\mathbf{R}) \\ P_2(\mathbf{R}) \end{pmatrix} = c\begin{pmatrix} 1 \\ 1 \end{pmatrix} + \frac{1}{4}\eta_0^{III}\begin{pmatrix} 1 \\ -1 \end{pmatrix} + \frac{1}{4}\left[\eta_1^{III}\begin{pmatrix} 1 \\ 1 \end{pmatrix} + \eta_2^{III}\begin{pmatrix} 1 \\ -1 \end{pmatrix}\right]\cos(\pi n_1)$, <br><br> $\begin{pmatrix} P_1(\mathbf{R}) \\ P_2(\mathbf{R}) \end{pmatrix} = c\begin{pmatrix} 1 \\ 1 \end{pmatrix} + \frac{1}{4}\eta_0^{III}\begin{pmatrix} 1 \\ -1 \end{pmatrix} + \frac{1}{4}\left[\eta_1^{III}\begin{pmatrix} 1 \\ -1 \end{pmatrix} + \eta_2^{III}\begin{pmatrix} 1 \\ 1 \end{pmatrix}\right]\cos(\pi n_2)$ |
| $C_7Me$ (Fig. 3) | $\begin{pmatrix} P_1(\mathbf{R}) \\ P_2(\mathbf{R}) \end{pmatrix} = c\begin{pmatrix} 1 \\ 1 \end{pmatrix} + \frac{1}{8}\eta_0^{III}\begin{pmatrix} 1 \\ -1 \end{pmatrix} +$ <br><br> $+ \frac{1}{8}\eta_1^{III}\left[\begin{pmatrix} 1 \\ 1 \end{pmatrix}\cos(\pi n_1) + \begin{pmatrix} 1 \\ -1 \end{pmatrix}\cos(\pi n_2) + \begin{pmatrix} 1 \\ 1 \end{pmatrix}\cos(\pi(n_1 - n_2))\right] +$ <br><br> $+ \frac{1}{8}\eta_2^{III}\left[\begin{pmatrix} 1 \\ -1 \end{pmatrix}\cos(\pi n_1) + \begin{pmatrix} 1 \\ 1 \end{pmatrix}\cos(\pi n_2) + \begin{pmatrix} 1 \\ -1 \end{pmatrix}\cos(\pi(n_1 - n_2))\right]$ |
| $CMe$ (Fig. 4a) | $\begin{pmatrix} P_1(\mathbf{R}) \\ P_2(\mathbf{R}) \end{pmatrix} = c\begin{pmatrix} 1 \\ 1 \end{pmatrix} + \frac{1}{2}\eta_1^I\begin{pmatrix} 1 \\ 1 \end{pmatrix}\cos(\pi(n_1 - n_2))$, <br><br> $\begin{pmatrix} P_1(\mathbf{R}) \\ P_2(\mathbf{R}) \end{pmatrix} = c\begin{pmatrix} 1 \\ 1 \end{pmatrix} + \frac{1}{2}\eta_1^I\begin{pmatrix} 1 \\ 1 \end{pmatrix}\cos(\pi n_1)$, $\quad \begin{pmatrix} P_1(\mathbf{R}) \\ P_2(\mathbf{R}) \end{pmatrix} = c\begin{pmatrix} 1 \\ 1 \end{pmatrix} + \frac{1}{2}\eta_1^I\begin{pmatrix} 1 \\ -1 \end{pmatrix}\cos(\pi n_2)$ |
| $CMe$ (Fig. 4b) | $\begin{pmatrix} P_1(\mathbf{R}) \\ P_2(\mathbf{R}) \end{pmatrix} = c\begin{pmatrix} 1 \\ 1 \end{pmatrix} + \frac{1}{2}\eta_2^I\begin{pmatrix} 1 \\ -1 \end{pmatrix}\cos(\pi(n_1 - n_2))$, <br><br> $\begin{pmatrix} P_1(\mathbf{R}) \\ P_2(\mathbf{R}) \end{pmatrix} = c\begin{pmatrix} 1 \\ 1 \end{pmatrix} + \frac{1}{2}\eta_2^I\begin{pmatrix} 1 \\ -1 \end{pmatrix}\cos(\pi n_1)$, $\quad \begin{pmatrix} P_1(\mathbf{R}) \\ P_2(\mathbf{R}) \end{pmatrix} = c\begin{pmatrix} 1 \\ 1 \end{pmatrix} + \frac{1}{2}\eta_2^I\begin{pmatrix} 1 \\ 1 \end{pmatrix}\cos(\pi n_2)$ |
| $CMe$ (Fig. 4c) | $\begin{pmatrix} P_1(\mathbf{R}) \\ P_2(\mathbf{R}) \end{pmatrix} = c\begin{pmatrix} 1 \\ 1 \end{pmatrix} + \frac{1}{2}\eta_0^I\begin{pmatrix} 1 \\ -1 \end{pmatrix}$ |

Table 2. Kinetic equations for LRO parameter(s) of different graphene-based (super)structures.

| (Super)structure | Differential kinetic equation(s) |
|---|---|
| $C_3Me$ (Fig. 2a) | $\dfrac{d\eta_2^I}{dt^*} = (1-c)\left[\dfrac{\eta_2^I}{T^*} - \ln\dfrac{(c+3\eta_2^I/4)(1-c+\eta_2^I/4)}{(1-c-3\eta_2^I/4)(c-\eta_2^I/4)}\right]$ |
| $C_3Me$ (Fig. 2b) | $\dfrac{d\eta_1^{II}}{dt^*} = c(1-c)\left[\dfrac{\eta_1^{II}}{T^*} - \ln\dfrac{(c+\eta_1^{II}/2+\eta_2^{II}/4)(1-c+\eta_1^{II}/2-\eta_2^{II}/4)}{(c-\eta_1^{II}/2+\eta_2^{II}/4)(1-c-\eta_1^{II}/2-\eta_2^{II}/4)}\right]$, $\dfrac{d\eta_2^{II}}{dt^*} = c(1-c)\left[\dfrac{\eta_2^{II}}{T^*} - \ln\dfrac{(c+\eta_1^{II}/2+\eta_2^{II}/4)(c-\eta_1^{II}/2+\eta_2^{II}/4)(1-c+\eta_2^{II}/4)^2}{(1-c-\eta_1^{II}/2-\eta_2^{II}/4)(1-c+\eta_1^{II}/2-\eta_2^{II}/4)(c-\eta_2^{II}/4)^2}\right]$ |
| $C_3Me$, (Fig. 2c) | $\dfrac{d\eta_0^{III}}{dt^*} = c(1-c)\left[\dfrac{\eta_0^{III}}{T^*} - \ln\left(\dfrac{(c+(\eta_0^{III}+\eta_1^{III}+\eta_2^{III})/4)(c+(\eta_0^{III}-\eta_1^{III}-\eta_2^{III})/4)}{(c+(-\eta_0^{III}+\eta_1^{III}-\eta_2^{III})/4)(c+(-\eta_0^{III}-\eta_1^{III}+\eta_2^{III})/4)}\right.\right.$ $\left.\left.\times\dfrac{(1-c-(-\eta_0^{III}+\eta_1^{III}-\eta_2^{III})/4)(1-c-(-\eta_0^{III}-\eta_1^{III}+\eta_2^{III})/4)}{(1-c-(\eta_0^{III}+\eta_1^{III}+\eta_2^{III})/4)(1-c-(\eta_0^{III}-\eta_1^{III}-\eta_2^{III})/4)}\right)\right]$, $\dfrac{d\eta_1^{III}}{dt^*} = c(1-c)\left[\dfrac{\eta_1^{III}}{T^*} - \ln\left(\dfrac{(c+(\eta_0^{III}+\eta_1^{III}+\eta_2^{III})/4)(c+(-\eta_0^{III}+\eta_1^{III}-\eta_2^{III})/4)}{(c+(\eta_0^{III}-\eta_1^{III}-\eta_2^{III})/4)(c+(-\eta_0^{III}-\eta_1^{III}+\eta_2^{III})/4)}\right.\right.$ $\left.\left.\times\dfrac{(1-c-(\eta_0^{III}-\eta_1^{III}-\eta_2^{III})/4)(1-c-(-\eta_0^{III}-\eta_1^{III}+\eta_2^{III})/4)}{(1-c-(\eta_0^{III}+\eta_1^{III}+\eta_2^{III})/4)(1-c-(-\eta_0^{III}+\eta_1^{III}-\eta_2^{III})/4)}\right)\right]$, $\dfrac{d\eta_2^{III}}{dt^*} = c(1-c)\left[\dfrac{\eta_2^{III}}{T^*} - \ln\left(\dfrac{(c+(\eta_0^{III}+\eta_1^{III}+\eta_2^{III})/4)(c+(-\eta_0^{III}-\eta_1^{III}+\eta_2^{III})/4)}{(c+(\eta_0^{III}-\eta_1^{III}-\eta_2^{III})/4)(c+(-\eta_0^{III}+\eta_1^{III}-\eta_2^{III})/4)}\right.\right.$ $\left.\left.\times\dfrac{(1-c-(\eta_0^{III}-\eta_1^{III}-\eta_2^{III})/4)(1-c-(-\eta_0^{III}+\eta_1^{III}-\eta_2^{III})/4)}{(1-c-(\eta_0^{III}+\eta_1^{III}+\eta_2^{III})/4)(1-c-(-\eta_0^{III}-\eta_1^{III}+\eta_2^{III})/4)}\right)\right]$ |
| $C_7Me$ (Fig. 3) | $\dfrac{d\eta_0^{III}}{dt^*} = c(1-c)\left[\dfrac{\eta_0^{III}}{T^*} - \ln\left(\dfrac{(c+(\eta_0^{III}+3\eta_1^{III}+3\eta_2^{III})/8)(c+(\eta_0^{III}-\eta_1^{III}-\eta_2^{III})/8)^3}{(1-c-(\eta_0^{III}+3\eta_1^{III}+3\eta_2^{III})/8)(1-c-(\eta_0^{III}-\eta_1^{III}-\eta_2^{III})/8)^3}\right.\right.$ $\left.\left.\times\dfrac{(1-c-(-\eta_0^{III}-3\eta_1^{III}+3\eta_2^{III})/8)(1-c-(-\eta_0^{III}+\eta_1^{III}-\eta_2^{III})/8)^3}{(c+(-\eta_0^{III}-3\eta_1^{III}+3\eta_2^{III})/8)(c+(-\eta_0^{III}+\eta_1^{III}-\eta_2^{III})/8)^3}\right)\right]$, $\dfrac{d\eta_1^{III}}{dt^*} = c(1-c)\left[\dfrac{\eta_1^{III}}{T^*} - \ln\left(\dfrac{(c+(\eta_0^{III}+3\eta_1^{III}+3\eta_2^{III})/8)(c+(-\eta_0^{III}+\eta_1^{III}-\eta_2^{III})/8)}{(c+(-\eta_0^{III}-3\eta_1^{III}+3\eta_2^{III})/8)(c+(\eta_0^{III}-\eta_1^{III}-\eta_2^{III})/8)}\right.\right.$ $\left.\left.\times\dfrac{(1-c-(-\eta_0^{III}-3\eta_1^{III}+3\eta_2^{III})/8)(1-c-(\eta_0^{III}-\eta_1^{III}-\eta_2^{III})/8)}{(1-c-(\eta_0^{III}+3\eta_1^{III}+3\eta_2^{III})/8)(1-c-(-\eta_0^{III}+\eta_1^{III}-\eta_2^{III})/8)}\right)\right]$, $\dfrac{d\eta_2^{III}}{dt^*} = c(1-c)\left[\dfrac{\eta_2^{III}}{T^*} - \ln\left(\dfrac{(c+(\eta_0^{III}+3\eta_1^{III}+3\eta_2^{III})/8)(c+(-\eta_0^{III}-3\eta_1^{III}+3\eta_2^{III})/8)}{(c+(\eta_0^{III}-\eta_1^{III}-\eta_2^{III})/8)(c+(-\eta_0^{III}+\eta_1^{III}-\eta_2^{III})/8)}\right.\right.$ $\left.\left.\times\dfrac{(1-c-(\eta_0^{III}-\eta_1^{III}-\eta_2^{III})/8)(1-c-(-\eta_0^{III}+\eta_1^{III}-\eta_2^{III})/8)}{(1-c-(\eta_0^{III}+3\eta_1^{III}+3\eta_2^{III})/8)(1-c-(-\eta_0^{III}-3\eta_1^{III}+3\eta_2^{III})/8)}\right)\right]$ |

| | | |
|---|---|---|
| CMe (Fig. 4a) | $\dfrac{d\eta_1^I}{dt^*} = c(1-c)\left[\dfrac{\eta_1^I}{T^*} - \ln\dfrac{(c+\eta_1^I/2)(1-c+\eta_1^I/2)}{(c-\eta_1^I/2)(1-c-\eta_1^I/2)}\right]$ | |
| CMe (Fig. 4b) | $\dfrac{d\eta_2^I}{dt^*} = c(1-c)\left[\dfrac{\eta_2^I}{T^*} - \ln\dfrac{(c+\eta_2^I/2)(1-c+\eta_2^I/2)}{(c-\eta_2^I/2)(1-c-\eta_2^I/2)}\right]$ | |
| CMe (Fig. 4c) | $\dfrac{d\eta_0^I}{dt^*} = c(1-c)\left[\dfrac{\eta_0^I}{T^*} - \ln\dfrac{(c+\eta_0^I/2)(1-c+\eta_0^I/2)}{(c-\eta_0^I/2)(1-c-\eta_0^I/2)}\right]$ | |

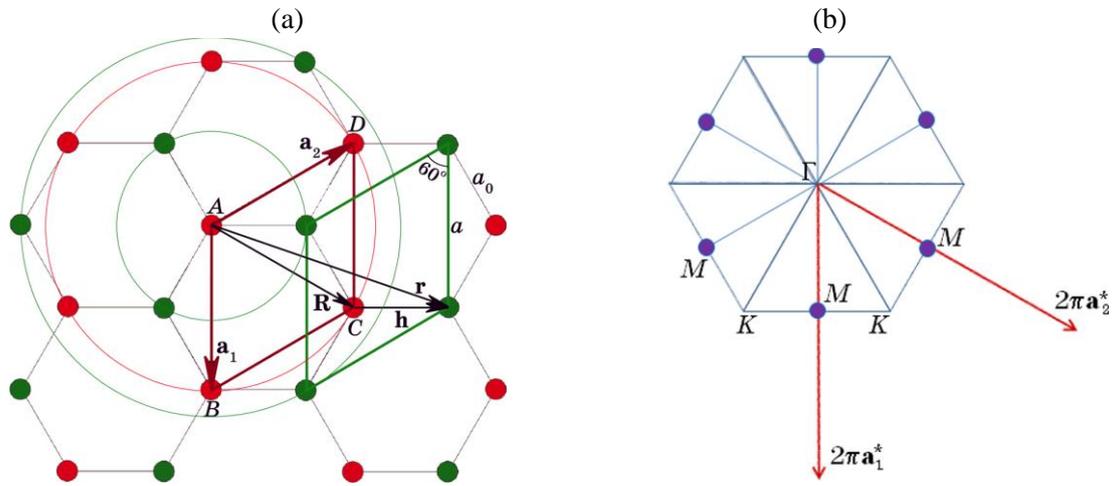

Fig. 1. The real-space lattice of graphene (a) and its reciprocal space (b). Here, (a) *ABCD* is a primitive unit cell, $\mathbf{a}_1$ and $\mathbf{a}_2$ are the basis translation vectors of the lattice, *a* is the lattice translation parameter, $a_0$ is a distance between the nearest-neighbour sites, circles denote the first three coordination shells with respect to the 'origin' (at site *A*) of the oblique coordinate system; (b) the first Brillouin zone ($\Gamma$, *M*, *K* are its high-symmetry points), $\mathbf{a}_1^*$ and $\mathbf{a}_2^*$ —the basis translation vectors of reciprocal lattice.

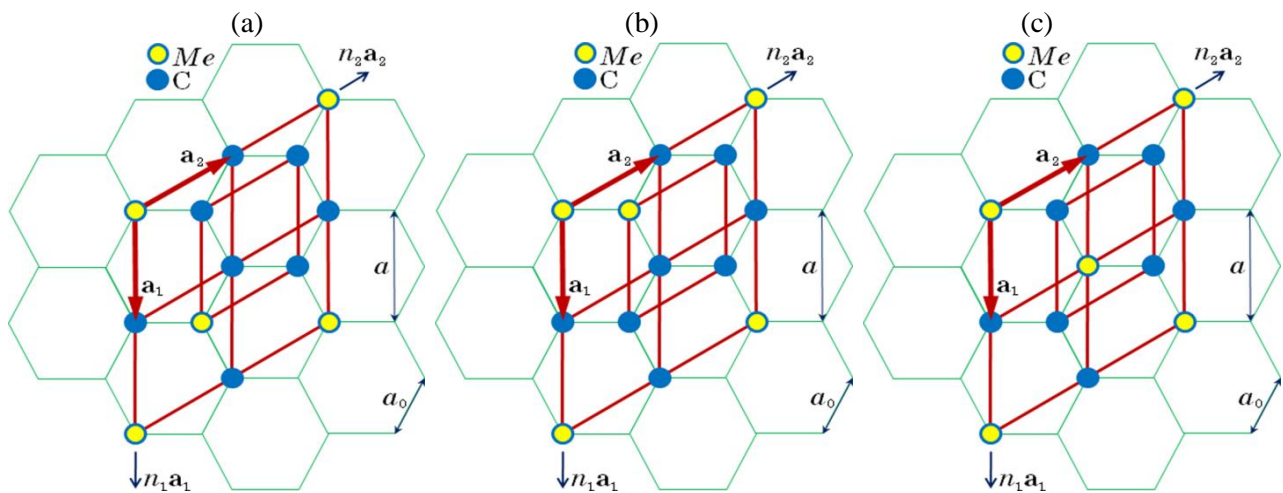

Fig. 2. Primitive unit cells of graphene-based $C_3Me$ superstructures described by one (a), two (b) or three (c) LRO parameters.

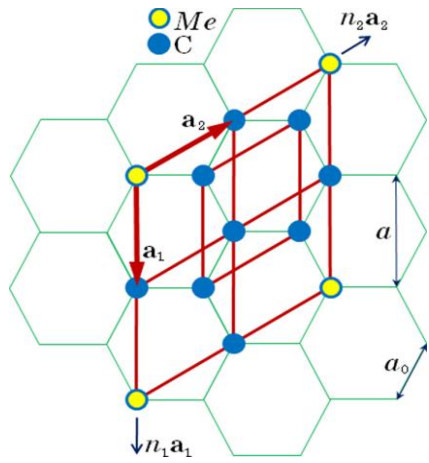

Fig. 3. Primitive unit cell of graphene-based C$_7$*Me*-superstructure described by three LRO parameters.

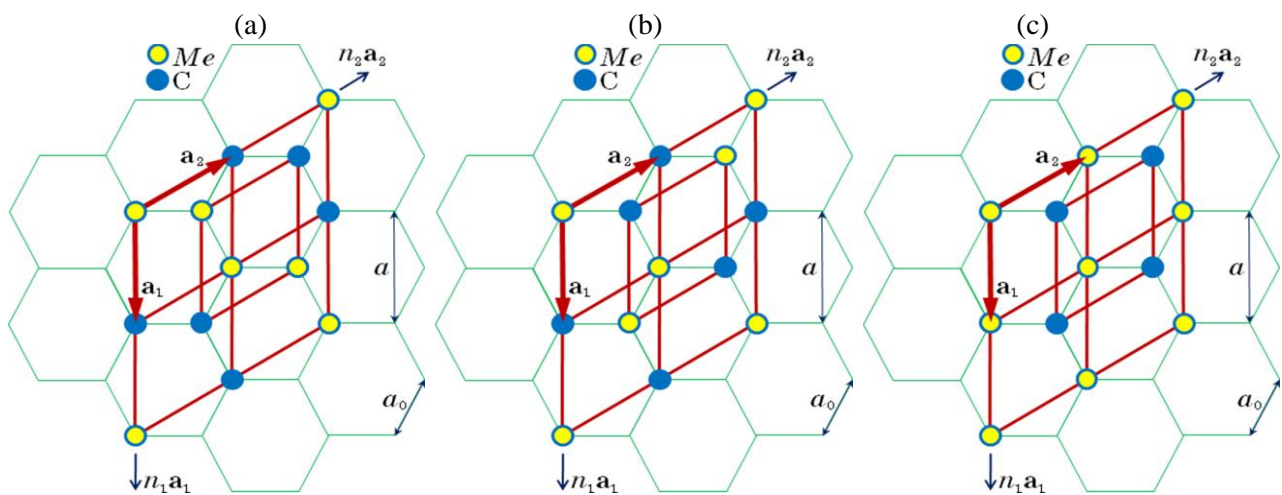

Fig. 4. Primitive unit cells of graphene-based C*Me*-superstructures described by one LRO parameter.

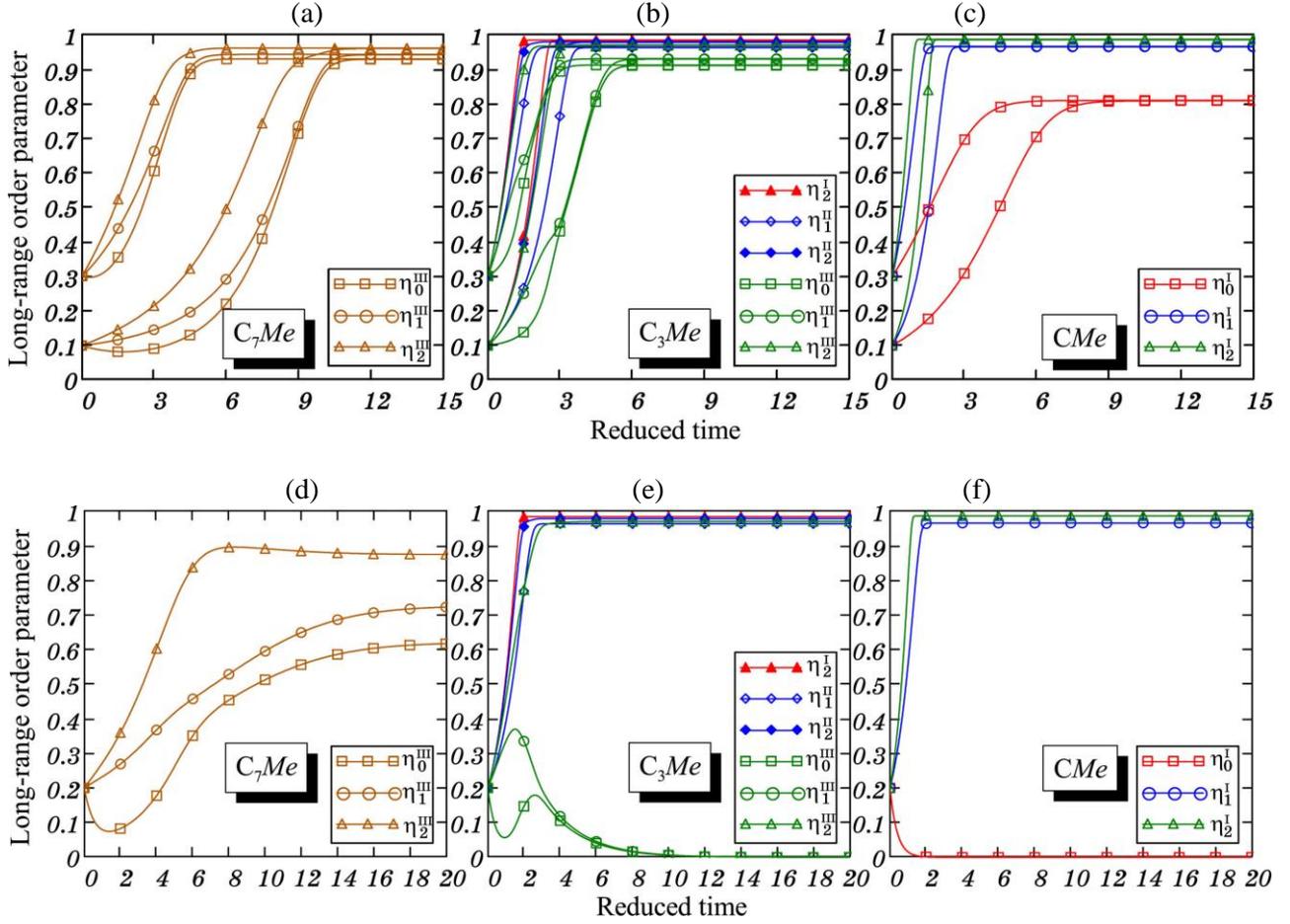

Fig. 5. The time evolution of the LRO parameter(s) for stoichiometric graphene-based structures at the fixed reduced temperature $T^* = k_B T/|\lambda_2(\mathbf{k})| = 0.1$ ($\lambda_2(\mathbf{k}) < 0$) and different initial conditions. Here, (a)–(c) $\lambda_2(\mathbf{0})/\lambda_2(\mathbf{k}) = 5/9$ and $\lambda_1(\mathbf{0})/\lambda_2(\mathbf{k}) = 5/6$; (d)–(f) $\lambda_2(\mathbf{0})/\lambda_2(\mathbf{k}) = -5/8$ and $\lambda_1(\mathbf{k})/\lambda_2(\mathbf{k}) = 5/6$.

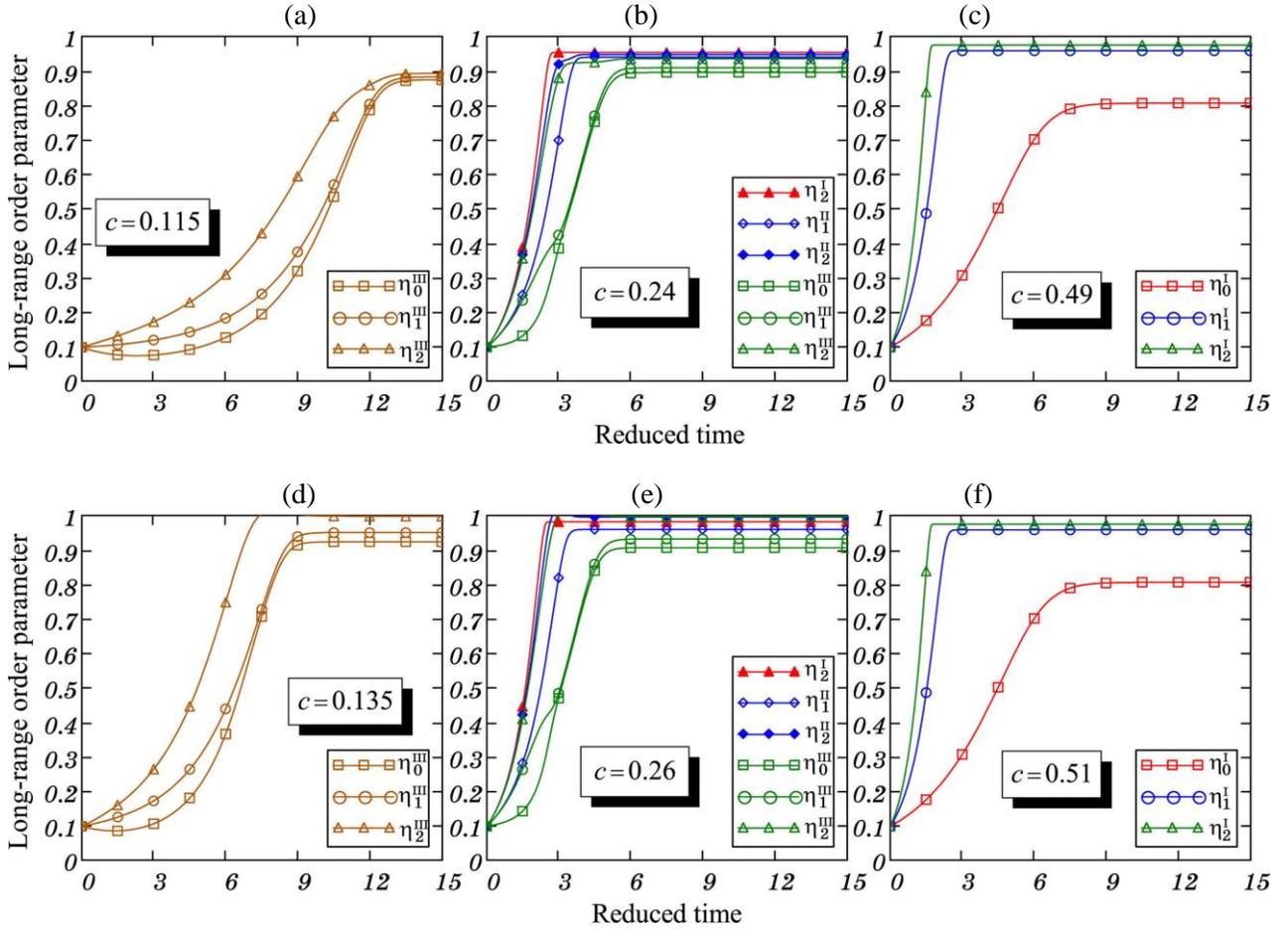

Fig. 6. The time dependence of LRO parameter(s) for nonstoichiometric graphene-based $C_{1-c}Me_c$ structures ($T^*$, $\lambda_2(\mathbf{0})/\lambda_2(\mathbf{k})$ and $\lambda_1(\mathbf{k})/\lambda_2(\mathbf{k})$ are the same as in Figs. 5a–5c).